\documentclass[prl,aps,superscriptaddress,preprint,floatfix,nofootinbib]{revtex4-1}

\usepackage{epsf}
\usepackage{graphicx}
\usepackage{sidecap}

\usepackage{color}

\def \beq {\begin{equation}}
\def \eeq {\end{equation}}
\begin{document}

\title{{Discovery of the topological surface state in a noncentrosymmetric superconductor BiPd}}


\author{Madhab~Neupane*}
\affiliation {Condensed Matter and Magnet Science Group, Los Alamos National Laboratory, Los Alamos, NM 87545, USA}

\author{Nasser~Alidoust*}\affiliation {Joseph Henry Laboratory and Department of Physics, Princeton University, Princeton, New Jersey 08544, USA}

\author{Su-Yang Xu}
\affiliation {Joseph Henry Laboratory and Department of Physics, Princeton University, Princeton, New Jersey 08544, USA}



\author{Ilya~Belopolski}
\affiliation {Joseph Henry Laboratory and Department of Physics, Princeton University, Princeton, New Jersey 08544, USA}

\author{Daniel S.~Sanchez}
\affiliation {Joseph Henry Laboratory and Department of Physics, Princeton University, Princeton, New Jersey 08544, USA}

\author{Tay-Rong Chang}
\affiliation{Department of Physics, National Tsing Hua University, Hsinchu 30013, Taiwan}

\author{Horng-Tay Jeng}
\affiliation{Department of Physics, National Tsing Hua University, Hsinchu 30013, Taiwan}
\affiliation{Institute of Physics, Academia Sinica, Taipei 11529, Taiwan}

\author{Hsin~Lin}
\affiliation{Centre for Advanced 2D Materials and Graphene Research Centre,
National University of Singapore, Singapore 117546}
\affiliation{Department of Physics, National University of Singapore,
Singapore 117542}


\author{Arun~Bansil}
\affiliation {Department of Physics, Northeastern University,
Boston, Massachusetts 02115, USA}


\author{Dariusz Kaczorowski}
\affiliation {Institute of Low Temperature and Structure Research, Polish Academy of Sciences,
50-950 Wroclaw, Poland}

\author{M.~Zahid~Hasan}
\affiliation {Joseph Henry Laboratory and Department of Physics,
Princeton University, Princeton, New Jersey 08544, USA}

\author{Tomasz~Durakiewicz}
\affiliation {Condensed Matter and Magnet Science Group, Los Alamos National Laboratory, Los Alamos, NM 87545, USA}

\date{18 June, 2013}
\pacs{}
\begin{abstract}

\textbf{Recently, noncentrosymmetric superconductor BiPd has attracted considerable research
interest due to the possibility of being a topological superconductor. Here, we report a systematic high-resolution angle resolved photoemission spectroscopy (ARPES) study of the normal state electronic properties of BiPd. Our experimental results show the presence of  a surface state at higher binding energy with the location of Dirac point at around 700 meV below the Fermi level. The detailed photon energy and temperature dependent measurements complemented by our first-principles calculations provide further evidence for the presence of the topological surface state at high binding energy. 
The absence of topological surface states near the Fermi level negates the possibility of the topological superconducting behavior in the surface of this material.
Our first direct experimental discovery of a topological surface state in BiPd
provides novel information that will guide the future search for topological superconductivity in noncentrosymmetric materials.}



\end{abstract}
\date{\today}
\maketitle


Recently, noncentrosymmetric (NCS) superconductors (SCs) have attracted considerable
research interest due to the possibility that they host several exotic properties \cite{note, Bauer, Sigrist}. The
lack of inversion symmetry creates an asymmetric potential gradient, which may
split the electron bands by lifting the spin degeneracy, allowing hybrid-pairing of
spin-singlet and spin-triplet states within the same orbital channel \cite{Sigrist, PRB_2011, PRB_2014, arXiv_2014}. Furthermore, with the advent of topological insulators \cite{Hasan, Hsieh, SCZhang, Hasan_review_2, Neupane, Superconducting_TI, Neupane_2, Xia}, it was
recently proposed that NCS SCs with a strong spin-orbit coupling are potential candidates
for realizing topological superconductivity \cite{Heusler_2010, CKLu, Hsin_Heusler,Ludwig_classification}, which hold promise for hosting protected Majorana surface states \cite{Majorana, Superconducting_TI, Vishwanath_PRL2011}.
But so far, direct evidence of the topological surface states in noncentrosymmetric materials is still lacking. Such convincing evidence may be obtained by angle-resolved photoemission, which provides an energy and momentum-resolved probe of electronic structure.


The noncentrosymmetric superconductor BiPd provides a platform to study the interplay
of spin-orbit coupling (SOC) effects with superconductivity. This compound undergoes a
structural transition from $\beta$ - BiPd (orthorhombic ) to $\alpha$ - BiPd (monoclinic) at 210 $^{\circ}$C and
then becomes superconducting at transition temperature ($T_c$) $\sim$ 3.7 K \cite{PRB_2011, PRB_2014}. In comparison to many other NCS
SCs, BiPd is a weakly correlated compound possessing a heavy atom, Bi. Measurements
by point-contact spectroscopy \cite{tunelling} and nuclear quadrupole resonance (NQR) \cite{NMR} indicate a
complex gap structure in BiPd, which might be caused by the lack of inversion symmetry
\cite{PRB_2011, PRB_2014}. Recent scanning tunneling microscopy (STM) measurements reveal that the
superconducting state of BiPd appears to be topologically trivial, consistent with
Bardeen-Cooper-Schrieffer theory with an $s$-wave order parameter \cite{arXiv_2014}. 
 Further momentum-resolved experimental evidence is highly
desirable in order to establish the presence of topological Dirac surface states, and to study the underlying microscopic mechanism of superconductivity.
A detailed systematic high-resolution angle-resolved photoemission spectroscopy (ARPES) study is needed to prove or disprove the topological nature of the surface states in the normal state of BiPd.
Such a characterization of the normal state is currently missing, but necessary and needed as a first step towards opening a discussion on the relationship between superconductivity and
the topological properties of noncentrosymmetric systems.

In this paper, we report the first experimental discovery of topological surface states in the noncentrosymmetric material  BiPd using ARPES. Our experimental results show the presence of  a surface state at higher binding energy, with the Dirac node located at around 700 meV below the Fermi level. Our detailed and systematic measurements provide the necessary evidence to demonstrate that this surface state is topological. Our results are further supported by our first-principle calculations. The absence of topological surface states near the Fermi level of BiPd negates the possibility that this system hosts
the topological superconductivity in the surface.
Our results provide a new direction for finding topological superconductivity in noncentrosymmetric materials.




\bigskip
\bigskip
\textbf{Results}

The crystal structure of BiPd at low temperatures ($< $ 210 $^{\circ}$C) has a monoclinic unit cell with $a= 5.63$ $\AA$,  $b= 10.66$ $\AA$, $c= 5.68$ $\AA$, $\alpha = \gamma =90^{\circ}$, and $\beta =101^{\circ}$ with the $b$ axis being its unique axis (see Fig. 1a and Ref. \cite{PRB_2011}).
Detailed characterization of the single crystals used in our study (see Supplementary Information) indicated their high quality. They exhibit a simple metallic behavior in the normal state and a sharp superconducting transition at  $T_c$ $\sim$ 3.7 K. The low-temperature magnetic susceptibility data is shown in Fig. 1b, while a picture of the single crystal measured is displayed in the inset of Fig. 1d.
A schematic bulk Brillouin zone is shown in Fig. 1c, where the projected surface along (010) is also illustrated. 
In order to experimentally identify its electronic structure, we systematically study the electronic
structure of BiPd on the cleaved (010) surface. Figure 1d shows
momentum-integrated ARPES spectral intensity over a wide
energy window. Sharp ARPES intensity peaks at binding energies
$E_B$ $\sim$  23 eV and 26 eV corresponding to the bismuth 5$d_{3/2}$ and
 5$d_{5/2}$ energy levels are observed.

We study the overall electronic structure of BiPd. Figure 2a shows an ARPES dispersion map in a 1.3 eV binding energy window, where several dispersive bands within the valence band are identified. 
Moreover, several crossing or metallic bands in the vicinity of  the Fermi level are observed.  Remarkably, a nearly linearly dispersive Dirac cone like state is observed at the Brillouin zone centre, showing a Dirac node located at a binding energy of $E_B$ $\sim$ 700 meV. The Dirac like state can be observed in the region of blue rectangle in the Fig. 2a at  binding energy region of 500 meV$-$900 meV.  At the Fermi level, only the metallic bands but no other Dirac like linearly dispersive bands are observed. On the other hand, the linearly dispersive Dirac like bands are found to be at the region of higher binding energies.

Fig. 2b shows the Fermi surface and constant energy contours plots. In the vicinity of the Fermi level, many metallic bands are observed (see Fig. 2b top left panel). 
We also study the ARPES measured constant energy contour maps
(Fig. 2b). At the Fermi level, the constant energy contour
consists of many metallic pockets. With increasing
binding energy at about 600 meV, the circular pocket formed by the upper Dirac cone is observed.
Upon increasing the binding energy, the size of the pocket decreases and eventually
shrinks to a point (the Dirac point) near $E_B$ $\sim$ 700 meV. Further increasing the binding energy, the nearly circular pocket formed by the lower Dirac cone is observed (see for $E_B$ $\sim$ 750 meV in Fig. 2b).

Figure 3 shows the energy-momentum cuts measured with varying photon energies from 30 eV to 58 eV with a 4 eV energy step (see Supplementary Information for ARPES spectra measured at low photon energy). Clear $E$-$k$ dispersion of bulk bands is observed. Remarkably, the dispersion of the linearly dispersive states at high-binding energy (500 meV to 900 meV) is found to be unchanged with respect to the varying photon energy, supporting the two-dimensional (2D) nature of this state.
It is important to note that the Dirac like 2D states are not found to be perfectly linear in energy-momentum axis. Furthermore, it is important to recall that in real materials such as pure Bi or graphene or topological insulators, the Dirac cones are never perfectly linear over a large energy window yet they can be approximated as linear within a narrow energy window around the Dirac point. 
This linear part represents the massless dispersion, in contrast to the large effective mass of conventional band electrons in other materials.


In order to test the robustness of the surface state observed in BiPd, we have performed a systematic temperature dependent measurement as shown in Fig. 4a.  Upon raising the temperature, the Dirac like surface states survive even at room temperature, which establishes that  the Dirac surface states are robust to the rise of temperature,
see Fig. 4a. Furthermore, the observed Dirac like states are found to be robust against thermal cycling, since lowering the temperature back down to 20 K results in the similar
spectra with the strong presence of Dirac like state features (see rightmost panel of Fig. 4a with the note of Re\_20K).

In order to better understand the electronic structure observed with ARPES, we perform first-principles
calculations on the bulk band structure (see Supplementary Information) and slab calculation of BiPd using the generalized gradient approximation (GGA) plus spin-orbit coupling (SOC) method (see Fig. 4b). Our slab calculations show that the surface state is predicted to be at $\Gamma$ point with the location of the Dirac node at around 0.5 eV and 0.6 eV below the Fermi level, one of which comes from the top surface and the another comes from the bottom surface. The Dirac node of the surface state coming from the top and bottom surfaces are located at different binding energy (not degenerate), which comes from the fact that BiPd lacks inversion symmetry. Calculations show that the Dirac surface state arises from the band inversion between Bi 6$p$ and Pd 4$p$ bands. Most importantly, the topological insulator property in BiPd originates from two valence bands inversion, which is very distinct with ordinary topological insulators \cite{Hasan, SCZhang, Xia},  where valence and conduction bands are involved during the band inversion process.
Our calculations qualitatively agree with our experimental results and calculations of Ref. \cite{arXiv_2014}. Our calculations further show a helical spin polarization of the surface state confirming its topological origin (see Supplementary Information). 
We note that the small deviation of the binding energy location of the Dirac point between experiment and theory may come from the natural doping behavior of BiPd, with
similar behavior also found in Bi$_2$Se$_3$ \cite{Xia}.







%

\bigskip

\bigskip
\textbf{Discussion}

The experimental realization of the topological insulator phase in a noncentrosymmetric crystal structure is an object of intense  research. Such a system may be utilized in testing several proposed exotic phenomena such as crystalline-surface-dependent topological electronic states, pyroelectricity, and natural topological $p$-$n$ junctions \cite{theory}. 
Recently, first principles calculations predicted III-Bi
to be an inversion asymmetric topological insulator with large band gap possessing intrinsic topologically protected edge states and forming quantum spin Hall systems \cite{theory} but these have not yet been realized experimentally. At the same time, 
the proposal of topological insulating nature in an inversion asymmetric compound BiTeCl still remains under debate \cite{YLChen_BiTeCl, Hugo_BiTeCl}. 
Furthermore, the small bulk band gaps of the realized inversion asymmetric topological insulators severely limit the manipulation and control of the topological surface states.
Most importantly, our discovery provide a unique example of topological surface state in a metallic compound.
Our detailed experimental and calculation results discover the first topological surface state in a noncentrosymmetric material and provide a new avenue to realize the properties proposed in non-centrosymmetric systems with topological surface states.



BiPd is a superconductor below $T_c$ $\sim$ 3.7 K. Since the topological surface states is located at high binding energy, it negates the topological superconductivity behavior in the surface at its native Fermi level. 
However, by electrical gating or surface deposition, the Fermi level can be tuned near the Dirac surface state, which provides an  opportunity  to  realize the topological superconductivity in this noncentrosymmetic materials.  
Our discovery of the topological surface state in the BiPd sample points towards a new direction in
the search for non-centrosymmetric TIs, promising novel applications based on topological states.
\newline
\newline
\textbf{Methods}
\newline
\textbf{Crystal growth and characterization.}
Single crystals of BiPd were grown by a modified Bridgman method as described elsewhere \cite{PRB_2011}. 
The crystals were characterized by means of X-ray diffraction, energy dispersive X-ray spectroscopy, magnetic susceptibility, electrical resistivity and heat capacity measurements, using standard commercial equipment.   
\newline
\newline
\textbf{Electronic structure measurements.}
Synchrotron-based ARPES measurements of the electronic structure were performed at the Advanced Light Source (ALS), Berkeley at Beamline 10.0.1 and Stanford Synchrotron Radiation Lightsource (SSRL) at Beamline 5-4 both equipped with a high efficiency R4000 electron analyzer. The energy resolution was set to be better than 20 meV for the measurements with the synchrotron beamline. The angular resolution was set to be better than  0.2$^{\circ}$ for all synchrotron measurements. 
Samples were cleaved in situ and measured at 10-80 K in a vacuum better than 10$^{-10}$ torr. They were found to be very stable and without degradation for the typical measurement
period of 20 hours.
\newline
\newline
\textbf{First-principles calculations.}
The first-principles calculations were based on the generalized gradient approximation (GGA) \cite{GGA} using the projector augmented-wave method \cite{PAW} as implemented in the VASP package \cite{VASP,VASP_1}. The experimental crystallographic structure was used \cite{expt} for the calculations. The spin-orbit coupling was included self-consistently in the electronic structure calculations with a 6 $\times$ 4 $\times$ 5 Monkhorst-Pack $k$-mesh.
In order to simulate surface effects, we used 1 $\times$ 5 $\times$ 1 supercell for the (010) surface, with vacuum thickness larger than 20 \AA.




\bigskip
\bigskip
\bigskip
\hspace{0.5cm}
\textbf{Acknowledgements}
\newline

M.N. was supported by LANL LDRD Program. T.D. was supported by Department of Energy, Office of Basic Energy Sciences, Division of Materials Sciences, and by NSF IR/D program.
The work at Princeton and synchrotron X-ray-based measurements and the related theory at Northeastern University are supported by the Office of Basic Energy Sciences, US Department of Energy (grants DE-FG-02-05ER46200, AC03-76SF00098 and DE-FG02-07ER46352).
The use of Synchrotron Radiation Center (SRC) was supported by NSF DMR-0537588 under the external user agreement.
H.L. acknowledges the Singapore National Research Foundation for the support under NRF Award No. NRF-NRFF2013-03. 
T.R.C. and H.T.J. were supported by the National Science Council, Taiwan. We also thank NCHC, CINC-NTU, and NCTS, Taiwan for technical support.
We thank Sung-Kwan Mo and Makoto Hashimoto for beamline assistance at the LBNL and the SSRL.
 M.Z.H. acknowledges Visiting Scientist support from LBNL and additional support from DOE/BES and the A. P. Sloan Foundation.

\*Correspondence and requests for materials should be addressed to M.N. (Email: mneupane@lanl.gov).

\begin{figure*}
\centering
\includegraphics[width=16.5cm]{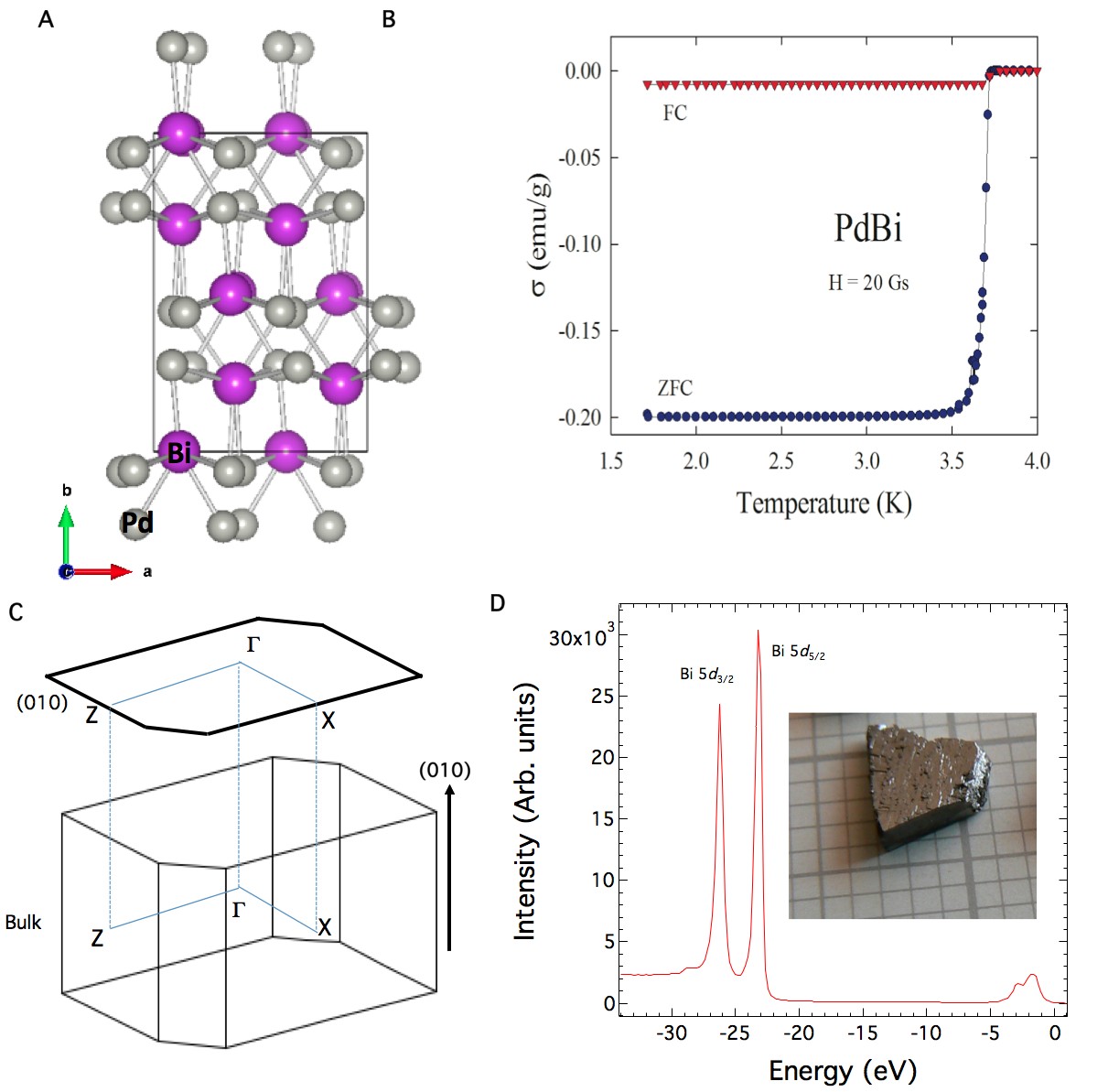}
\caption{\textbf{Crystal structure and sample characterization of BiPd}. 
\textbf{a,} Crystal structure of BiPd with the $b$ axis shown as its unique axis. It crystallizes in a monoclinic structure at low temperature.  \textbf{b,} The magnetic susceptibility as a function of temperatures showing a sharp superconducting transition temperature at $\sim$ 3.7 K in the field of 20 Gs. \textbf{c,} Schematic surface and bulk Brillouin zones are shown. High-symmetric points are also marked.
\textbf{d,} Core level spectroscopic measurement of BiPd showing sharp peaks of Bi 5$d$. The inset shows a photograph of the BiPd sample.}
\end{figure*}

\begin{figure*}
\centering
\includegraphics[width=16.5cm]{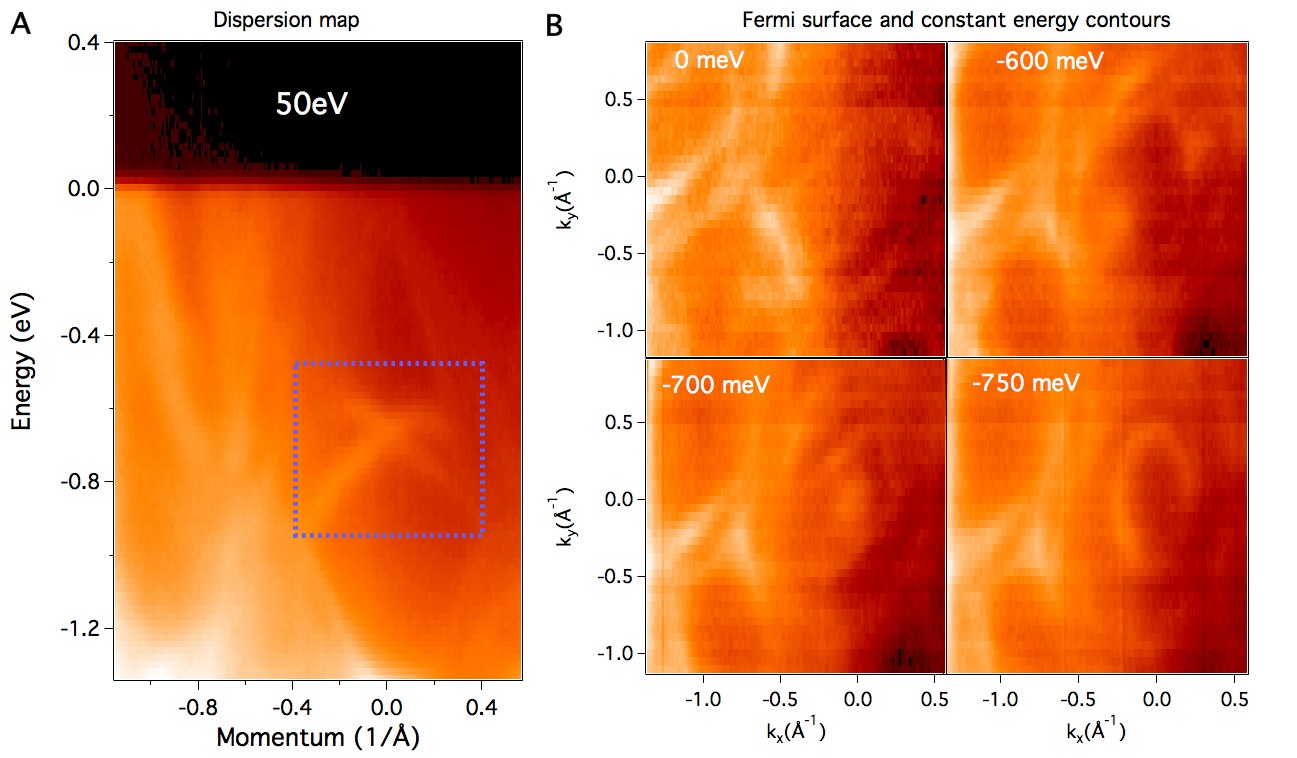}
\caption{\textbf{Dispersion map, Fermi surface and constant energy contours of BiPd.} \textbf{a,} Dispersion map of BiPd  along the zone centre obtained by using incident photon energy of 50 eV at a temperature of 10 K. The blue rectangle in the binding energy range of about  500 meV  to 900 meV shows the linearly dispersive states. \textbf{b,} Fermi surface map (with binding energy of 0 meV shown in  the top left panel) and constant energy contours. Constant energy contour at binding energy of 600 meV shows the intensity map in the region above the Dirac point. Similarly, constant energy contours at 700 meV and 750 meV show the intensity map at around the Dirac point and below the Dirac point, respectively. Data were collected at ALS BL 10.0.1}
\end{figure*}

\begin{figure*}
\centering
\includegraphics[width=16.5cm]{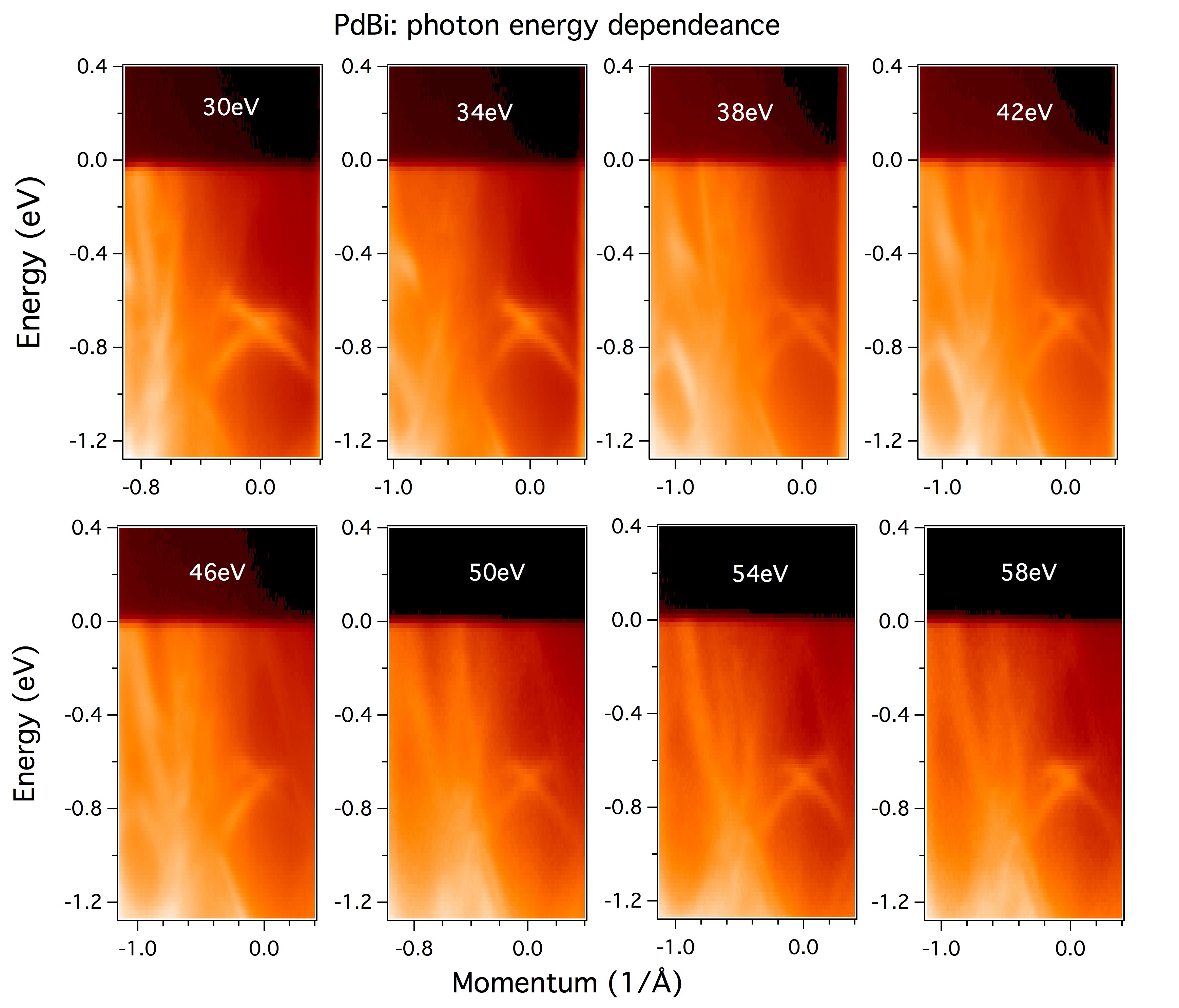}
\caption{\textbf{Photon energy dependent ARPES dispersion maps.}
The measured photon energies are noted on the plots. The linearly dispersive states at binding energy around 700 meV do not show any dispersion with photon energy, which indicate its two dimensional nature. These data were collected at ALS BL 10.0.1 at a temperature of 10 K.}
\end{figure*}



\begin{figure*}
\centering
\includegraphics[width=17.5cm]{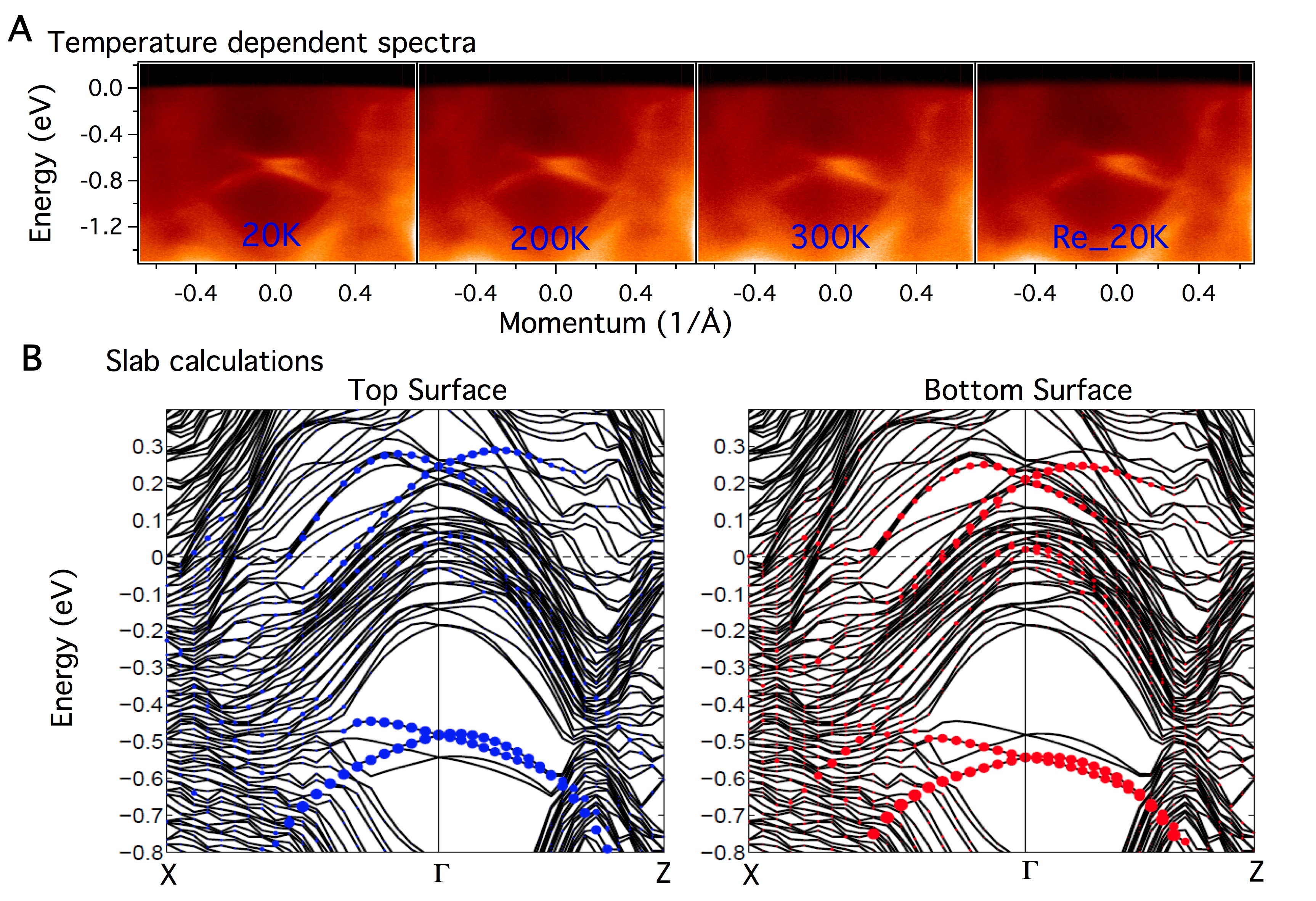}
\caption{\textbf{Temperature dependent ARPES measurements and first-principles calculations.} \textbf{a,} ARPES energy momentum dispersion maps measured using photon energy of 30eV along the ${\bar{X}}-{\bar{\Gamma}}-{\bar{X}}$ momentum space cut-direction with varying temperatures. The measured values of temperature are noted on the plots. The rightmost panel with the note of Re\_20K is the spectrum measured after thermal cycling (20K$\rightarrow$300K$\rightarrow$20K). These data were collected at SSRL BL5-4 with a photon energy of 30 eV. \textbf{b,} Slab calculations of BiPd for the top surface (left) and the bottom surface (right) along the high symmetry lines. The blue (red) dots represent surface states for the top (bottom) surface. Detailed of calculation method is given in Method Section and additional calculation plots are shown in Supplementary Information.}
\end{figure*}

%
%

\end{document}